\documentclass[journal]{IEEEtran}
\usepackage{cite}

\ifCLASSINFOpdf
   \usepackage[pdftex]{graphicx}
 \graphicspath{/Users/essox/Google_Drive/Poly/PhD/IPv6/}
\else
\fi
\usepackage{amsmath}
\usepackage{mathtools}

\usepackage{algorithm} 
\usepackage{algpseudocode}

\usepackage{color}

\ifCLASSOPTIONcompsoc
\usepackage[caption=false,font=normalsize,labelfont=sf,textfont=sf]{subfig}
\else
\usepackage[caption=false,font=footnotesize]{subfig}
\fi
\usepackage{url}

\usepackage[utf8]{inputenc}

\makeatletter
\newcommand{\algmargin}{\the\ALG@thistlm}
\makeatother
\newlength{\whilewidth}
\algdef{SE}[DOWHILE]{Do}{doWhile}{\algorithmicdo}[1]{\algorithmicwhile\ #1}%

\algdef{SE}[parIF]{parIf}{EndparIf}[1]
  {\parbox[t]{\dimexpr\linewidth - \algmargin}{%
     \hangindent\ifwidth\strut\algorithmicif\ #1\ \algorithmicdo\strut}}{\algorithmicend\ \algorithmicif}%
\algnewcommand{\parState}[1]{\State%
  \parbox[t]{\dimexpr\linewidth-\algmargin}{\strut #1\strut}}

\begin{document}
%
\title{SHIP: A Scalable High-performance IPv6 Lookup Algorithm that Exploits Prefix Characteristics}

%
%
%

\author{Thibaut Stimpfling, Normand Belanger, 
        J.M. Pierre~Langlois~\IEEEmembership{~Member~IEEE}
        and~Yvon~Savaria~\IEEEmembership{~Fellow~IEEE}\thanks{Thibaut Stimpfling, Normand Bélanger, J.M. Pierre Langlois, and Yvon Savaria are with École Polytechnique de Montréal (e-mail: \{thibaut.stimpfling, normand.belanger, pierre.langlois, yvon.savaria\}@polymtl.ca).}}

\maketitle

\begin{abstract}
Due to the emergence of new network applications, current IP lookup engines must support high-bandwidth, low lookup latency and the ongoing growth of IPv6 networks.
However, existing solutions are not designed to address jointly those three requirements.
This paper introduces SHIP, an IPv6 lookup algorithm that exploits prefix characteristics to build a two-level data structure designed to meet future application requirements. Using both prefix length distribution and prefix density, SHIP first clusters prefixes into groups sharing similar characteristics, then it builds a hybrid trie-tree for each prefix group. The compact and scalable data structure built can be stored in on-chip low-latency memories, and allows the traversal process to be parallelized and pipelined at each level in order to support high packet bandwidth. 


Evaluated on real and synthetic prefix tables holding up to 580 k IPv6 prefixes, SHIP has a logarithmic scaling factor in terms of the number of memory accesses, and a linear memory consumption scaling. Using the largest synthetic prefix table, simulations show that compared to other well-known approaches, SHIP uses at least 44\% less memory per prefix, while reducing the memory latency by 61\%. 

\end{abstract}


%
\IEEEpeerreviewmaketitle

\section{Introduction}
%
%
%
%

\IEEEPARstart{G}{lobal} IP traffic carried by networks is continuously growing, 
as around a zettabyte total traffic is expected for the whole of 2016, and it is envisioned to increase threefold between 2015 and 2019~\cite{cisco_forecast}. To handle this increasing Internet traffic, network link working groups have ratified the 100-gigabit Ethernet standard (IEEE P802.3ba), and are studying the 400-gigabit Ethernet standard (IEEE P802.3bs). As a result, network nodes have to process packets at those line rates which puts pressure on IP address lookup engines used in the routing process. Indeed, less than $6$ ns is available to determine the IP address lookup result for an IPv6 packet~\cite{scalable_ipv6_vk}. 

The IP lookup task consists of identifying the next hop information (NHI) to which a packet should be forwarded. The lookup process starts by extracting the destination IP field from the packet header, and then matching it against a list of entries stored in a lookup table, called the forwarding information base (FIB). Each entry in the lookup table represents a network defined by its prefix address. While a lookup key may match multiple entries in the FIB, only the longest prefix and its NHI are returned for result as IP lookup is based on the Longest Prefix Match (LPM)~\cite{high_performance_routers_switches}. 

IP lookup algorithms and architectures that have been tailored for IPv4 technology are not performing well with IPv6~\cite{scalable_ipv6_vk, mem_eff}, due to the fourfold increase in the number of bits in IPv6 addresses over IPv4. Thus, dedicated IPv6 lookup methods are needed to support upcoming IPv6 traffic.

IP lookup engines must be optimized for high bandwidth, low latency, and scalability for two reasons. First, due to the convergence of wired and mobile networks, many future applications that require a high bandwidth and a low latency, such as virtual reality, remote object manipulation, eHealth, autonomous driving, and the Internet of Things, will be carried on both wired and mobile networks~\cite{5g_whitepaper}. Second, the number of IPv6 networks is expected to grow, and so is the size of the IPv6 routing tables, as IPv6 technology is still being deployed in production networks~\cite{potaroo,ris_raw_data}. However, current solutions presented in the literature are not jointly addressing these three performance requirements. 

In this paper, we introduce SHIP: a Scalable and High Performance IPv6 lookup algorithm designed to meet current and future performance requirements. SHIP is built around the analysis of prefix characteristics. Two main contributions are presented: 1) two-level prefix grouping, that clusters prefixes in groups sharing common properties, based on the prefix length distribution and the prefix density, 2) a hybrid trie-tree tailored to handle prefix distribution variations. 
SHIP builds a compact and scalable data structure that is suitable for on-chip low-latency memories, and allows the traversal process to be parallelized and pipelined at each level in order to support high packet bandwidth. SHIP stores 580 k prefixes and the associated NHI using less than $5.9$ MB of memory, with a linear memory consumption scaling. SHIP achieves logarithmic latency scaling and requires in the worst case 10 memory accesses per lookup. For both metrics, SHIP outperforms known methods by over 44\% for the memory footprint, and by over 61\% for the memory latency.

The remainder of this paper is organized as follows. Section~\ref{sec:related_work} introduces common approaches used for IP lookup and Section~\ref{sec:overview} gives an overview of SHIP. Then, two-level prefix grouping is presented in Section~\ref{sec:two_level_prefix_grouping}, while the proposed hybrid trie-tree is covered in Section~\ref{sec:trie_tree_hybrid_data_structure}. Section~\ref{sec:method} introduces the method and metrics used for performance evaluation and Section~\ref{sec:results} presents the simulation results. Section~\ref{sec:disscusion} shows that SHIP fulfills the properties for hardware implementability, and compares SHIP performance with other methods. Lastly, we conclude the work by summarizing our main results in Section~\ref{sec:conclusion}.

\section{Related Work}\label{sec:related_work}
 
Many data structures have been proposed for the LPM operation applied to IP addresses. We can classify them in four main types: hash tables, Bloom filters, tries and trees. Those data structures are encoding prefixes that are loosely structured. First, not only prefix length distribution is highly nonuniform, but it also varies with the prefix table used. Second, for any given prefix length, prefix density ranges from sparse to very dense. Thus, each of the four main data structures type comes with a different tradeoff between time and storage complexity.

Interest for IP lookup with hash table is twofold. First, a hash function aims at distributing uniformly a large number of entries over a number of bins, independently of the structure of the data stored. Second, a hash table provides $O(1)$ lookup time and $O(N)$ space complexity. However, a pure hash based LPM solution can require up to one hash table per IP prefix length. An alternative to reduce the number of hash tables is to use prefix expansion~\cite{flashlook}, but it increases memory consumption. Two main types of hash functions can be selected to build a hash table: perfect or non-perfect hash functions. A Hash table built with a perfect hash functions offers a fixed time complexity that is independent from the prefixes used as no collision is generated. Nevertheless, a perfect hash function cannot handle dynamic prefix tables, making it unattractive for a pure hash based LPM solution. On the other hand, a non-perfect hash function leads to collisions and cannot provide a fixed time complexity. Extra-matching sequences are required with collisions that drastically decrease performance~\cite{flashtrie_conf,distributed_bloom_filters}. In addition, not only the number of collisions is determined after the creation of the hash table but it also depends on the prefix distribution characteristics. In order to reduce the number of collisions independently of the characteristics of the prefix table used, a method has been proposed that exploits multiple hash tables~\cite{flashlook,flashtrie_conf}. This method divides the prefix table into groups of prefixes, and selects a hash function such that it minimizes the number of collisions within each prefix group~\cite{flashlook,flashtrie_conf}. Still, the hash function selection for each prefix group requires to probe all the hash functions, making it unattractive for dynamic prefix tables. Finally, no scaling evaluation has been completed in recent publications~\cite{flashlook, flashtrie_journal} making it unclear whether the proposed hash-based data structures can address forthcoming challenges.

Low-memory footprint hashing schemes known as Bloom filters have also been covered in the literature~\cite{distributed_bloom_filters,bloom_filters}. Bloom filters are used to select a subgroup of prefixes that may match the input IP address. However, Bloom filters suffer from two drawbacks. First, by design, this data structure generates false positives independent of the configuration parameters used. Thus, a Bloom filter can improve the average lookup time, but it can also lead to poor performance in the worst case, as many sub-groups need to be matched. Second, the selection of a hash function that minimizes the number of false positives is highly dependent of the prefix distribution characteristics used. Hence, its complexity is similar to that of of a hash function that minimizes the number of collisions in a regular hash table.

Tree solutions based on binary search trees (BST) or generalized B-trees have also been explored in~\cite{mem_eff,scalable_ipv6_vk}. Such data structures are tailored to store loosely structured data such as prefixes, as their time complexity is independent from the prefix distribution characteristics. Indeed, BST and 2-3 Trees have a time complexity of respectively $log_2(N)$ and $log_3(N)$, with $N$ being the number of entries~\cite{scalable_ipv6_vk}. Nevertheless, such data structure provides a solution at the cost of a large memory consumption. Indeed, each node stores a full-size prefix, leading to memory waste. Hence, their memory footprint makes them unsuitable for the very large prefix tables that are anticipated in future networks.

At the other end of the tree spectrum, decision-trees (D-Trees) have been proposed in~\cite{hicuts,scalable_packet_classification} for the field of packet classification. D-Trees were found to offer a good tradeoff between memory footprint and the number of memory accesses. However, no work has been conducted yet on using this data structure for IPv6 lookup.

The trie data structure, also known as radix tree, has regained interest with tree bitmap~\cite{bitmap_tree}. Indeed, a $k$-bit trie requires $k/W$ memory accesses, but has very poor memory efficiency when built with unevenly distributed prefixes. A tree bitmap improves the memory efficiency over a multi-bit trie, independently of the prefix distribution characteristics, by using a bitmap to encode each level of a multi-bit trie. However, tree bitmaps cannot be used with large strides, as the node size grows exponentially with the stride size, leading to multiple wide memory accesses to read a single node. An improved tree bitmap, the PC-trie, is proposed for the FlashTrie architecture~\cite{flashtrie_journal}. A PC-Trie reduces the size of bitmap nodes using a multi-level leaf pushing method. This data structure is used jointly with a pre-processing hashing stage to reduce the total number of memory accesses. Nevertheless, the main shortcoming of the Flashtrie architecture lies in its pre-processing hashing module. First, similar to other hashing solutions, its performance highly depends on the distribution characteristics of the prefixes used. Second, the hashing module does not scale well with the number of prefixes used.

At the other end of the spectrum of algorithmic solutions, TCAMs have been proposed as a pure hardware solution, achieving $O(1)$ lookup time by matching the input key simultaneously against all prefixes, independently of their distribution characteristics. However, these solutions use a very large amount of hardware resources, leading to large power consumption and high cost, and making them unattractive for routers holding a large number of prefixes~\cite{a_tcam_based_distributed_parallel,flashtrie_journal}.

Recently, information-theoretic and compressed data structures have been applied to IP lookup, yielding very compact data structures, that can handle a very large number of prefixes~\cite{compressing_ip_forwarding_tables}. Even though this work is limited to IPv4 addresses, it is an important shift in terms of concepts. However, the hardware implementation of the architecture achieves 7 million lookups per second. In order to support a 100-Gbps bandwidth, this would require many lookup engines, leading to a memory consumption that is similar or higher than previous trie or tree algorithms~\cite{flashlook,flashtrie_conf,flashtrie_journal,scalable_ipv6_vk}.

In summary, the existing data structures are not exploiting the full potential of the prefix distribution characteristics.
In addition, none of the existing data structures were shown to optimize jointly the time complexity, the storage complexity, and the scalability. 

\section{SHIP Overview}\label{sec:overview}

SHIP consists of two procedures: the first one is used to build a two-level data structure, and the second one is used to traverse the two-level data structure. The procedure to build the data structure is called two-level prefix grouping, while the traversal procedure is called the lookup algorithm.
 
\begin{figure}[htbp]
\centering
  \includegraphics[scale=0.75]{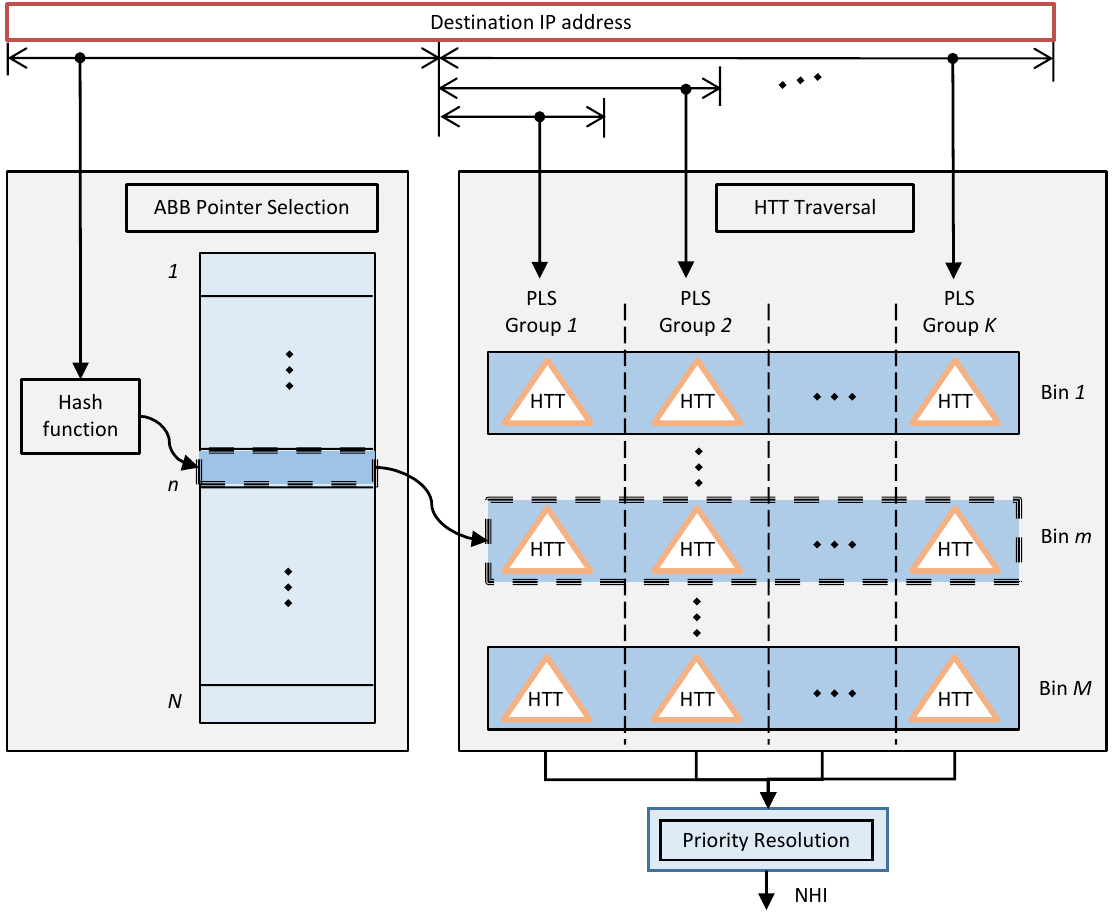}
  \caption{SHIP two-level data structure organization and its lookup process with $M$ address block bins and $K$ prefix length sorting groups.}
  \label{fig:global_lookup_seq}
\end{figure}

Two-level prefix grouping clusters prefixes upon their characteristics to build an efficient two-level data structure, presented in Fig~\ref{fig:global_lookup_seq}. At the first level, SHIP leverages the low density of the IPv6 prefix MSBs to divide prefixes into M address block bins (ABBs). A pointer to each ABB is stored in an N-entry hash table. At the second level, SHIP uses the uneven prefix length distribution to sort prefixes held in each ABB into K Prefix Length Sorted (PLS) groups. For each non-empty K$\cdot$M PLS groups, SHIP further exploits the prefix length distribution and the prefix density variation to encode prefixes into a hybrid trie-tree (HTT).

The lookup algorithm, which identifies the NHI associated to the longest prefix matched, is presented in Fig~\ref{fig:global_lookup_seq}. First, the MSBs of the destination IP address are hashed to select an ABB pointer stored in an $N$ entry hash table. The selected ABB pointer in this figure is held in the $n$-th entry of the hash table, represented with a dashed rectangle. This pointer identifies bin $m$, represented with a dashed rectangle. Second, the HTTs associated to each PLS group of the $m$-th bin, are traversed in parallel, using portions of the destination IP address. Each HTT can output a NHI, if a match occurs with its associated portion of the destination IP address. Thus, up to $K$ HHT results can occur, and a priority resolution module is used to select the NHI associated to the longest prefix.

In the following section, we present the two-level prefix grouping procedure.

\section{Two-level Prefix Grouping}\label{sec:two_level_prefix_grouping}

This section introduces two-level prefix grouping, that clusters and sort prefixes into groups, and then builds the two-level data structure. First, prefixes are binned and the first level of the two-level data structure is built with the address block binning method. Second, inside each bin, prefixes are sorted into groups, and then the HTTs are built with the prefix length sorting method.

 \subsection{Address block binning} 
 
The proposed address block binning method exploits both the structure of IP addresses and the low density of the IPv6 prefix MSBs to cluster the prefixes into bins, and then build the hash table used at the first level of SHIP data structure.

IPv6 addresses are structured into IP address blocks, managed by the Internet Assigned Numbers Authority (IANA) that assigns blocks of IPv6 addresses ranging in size from $/16$ to $/23$, that are then further divided into smaller address blocks. However, the prefix density on the first 23 bits is low, and the prefix distribution is  sparse~\cite{global_unicast_address_assig}. Therefore, the address block binning method bins prefixes based on their first 23 bits. Before prefixes are clustered, all prefixes with a prefix length that is less than $/23$ are converted into $/23$. The pseudo-code used for this binning method is presented in Algorithm~\ref{alg:build_address_block_binning}. For each prefix held in the prefix table, this method checks whether a bin already exists for the first $23$ bits. If none exists, a new bin is created, and the prefix is added to the new bin. Otherwise, the prefix is simply added to the existing bin. 
This method only keeps track of the $M$ created bins. The prefixes held in each bin are further grouped by the prefix length sorting method that is presented next.

Address block binning utilizes a perfect hash function~\cite{perfect_hashing_function} to store a pointer to each valid bin. Let $N$ be the size of the hash table used. The first 23 MSBs of the IP address represents the key of the perfect hash table. A perfect hashing function is chosen for three reasons. First, the number of valid keys on the first 23 bits is relatively small compared to the number of bits used, making hashing attractive. Second, a perfect hashing function is favoured when the data hashed is static, which is the case here, for the first 23 bits only, because it represents blocks of addresses allocated to regional internet registries that are unlikely to be updated on a short time scale. Finally, no resolution module is required because no collisions are generated.

\begin{algorithm}
\caption{Building the Address Block binning data structure}
\label{alg:build_address_block_binning}
\begin{algorithmic}[1]
\renewcommand{\algorithmicrequire}{\textbf{Input: }}
 \renewcommand{\algorithmicensure}{\textbf{Output: }}
 \Require Prefix table 
 \Ensure Address Block binning data structure 
\For{each prefix held in the prefix table}
\State {Extract its 23 MSBs}
\If {no bin already exists for the extracted bits}
\parState{Create a bin that is associated to the value of the extracted bits}
\State{Add the current prefix to the bin}
\Else 
\parState {Select the bin that is associated to the value of the extracted bits}
\State{Add the current prefix to the selected bin}
\EndIf
\EndFor
\State {Build a hash table that stores a pointer to each address block bin, with the 23 MSBs of each created bin as a key. Empty entries are holding invalid pointers.}
\State \textbf{return}  {the hash table}
\end{algorithmic}
\end{algorithm}

The lookup procedure is presented in Algorithm~\ref{alg:lookup_address_block_binning}. It uses the 23 MSBs of the destination IP address as a key for the hash table and returns a pointer to an address block bin. If no valid pointer exists in the hash table for the key, then a null pointer is returned. 

\begin{algorithm}
\caption{Lookup in the Address Block binning data structure}
\label{alg:lookup_address_block_binning}
\begin{algorithmic}[1]
\renewcommand{\algorithmicrequire}{\textbf{Input: }}
 \renewcommand{\algorithmicensure}{\textbf{Output: }}
 \Require{Address Block binning hash table, destination IP Address}
 \Ensure{Pointer to an address block bin}
\State{Extract the 23 MSBs of the destination IP address}
\State {Hash the extracted 23 MSBs}
\If{the hash table entry pointed by the hashed key holds a valid pointer}
\State{ \textbf{return} {pointer to the address block bin}}
\Else
\State{ \textbf{return}  {null pointer}}
\EndIf
\end{algorithmic}
\end{algorithm}

The perfect hash table created with the address block binning method is an efficient data structure to perform a lookup on the first 23 MSBs of the IP address. 
However, within the ABBs the prefix length distribution can be highly uneven, which degrades the performance of the hybrid trie-trees at the second level. Therefore, the prefix length sorting method, described next, is proposed to address that problem.

 \subsection{Prefix length sorting}

Prefix length sorting (PLS) aims at reducing the impact of the uneven prefix length distribution on the number of overlaps between prefixes held in each address block bin. By reducing the number of prefix overlaps, the performance of the HTTs is improved, as it will be shown later. The PLS method sorts the prefixes held in each address block bin by their length, into $K$ PLS groups that cover disjoints prefix length ranges. Each range consists of contiguous prefix lengths that are associated to a large number of prefixes with respect to the prefix table size. For each PLS group, a hybrid trie-tree is built.

The number of PLS groups, $K$, is chosen to maximize the HTT's performance. As will be shown experimentally in section~\ref{sec:results}, beyond a threshold value, increasing the value of $K$ does not further improve performance. The prefix length range selection is based on the prefix length distribution and it is guided by two principles. First, to minimize prefix overlap, when a prefix length covers a large percentage of the total number of prefixes, this prefix length must be used as an upper bound of the considered group. Second, prefix lengths included in a group are selected such that group sizes are as balanced as possible.

\begin{figure}
\centering
\includegraphics[width=3.5in]{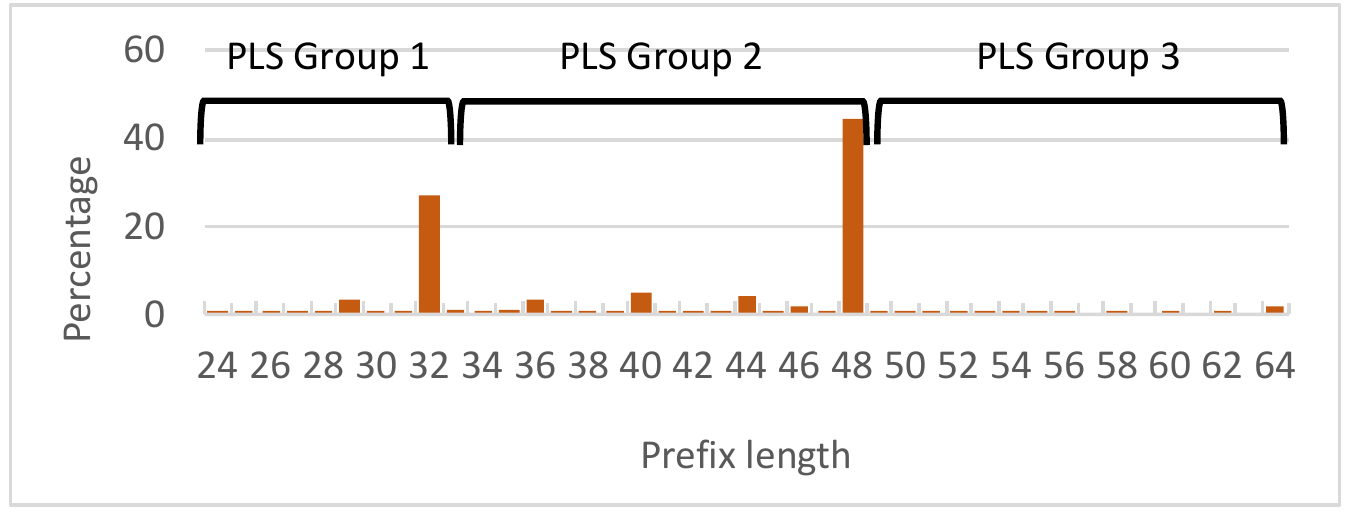}
\caption{The uneven prefix length distribution of real a prefix table used by the PLS method to create 3 PLS groups.}
\label{fig:prefix_distribution}
\end{figure}

To illustrate those two principles, an analysis of prefix length distribution using a real prefix table is presented in Fig.~\ref{fig:prefix_distribution}. The prefix table extracted from~\cite{ris_raw_data} holds approximately 25 k prefixes. The first $23$ prefix lengths are omitted in Fig.~\ref{fig:prefix_distribution}, as the address block binning method already bins prefixes on their $23$ MSBs. It can be observed in Fig.~\ref{fig:prefix_distribution} that the prefix lengths with the largest cardinality are $/32$ and $/48$ for this example. Applying the two principles of prefix length sorting to this example, the first group covers prefix lengths from $/24$ to $/32$, and the second group covers the second peak, from $/33$ to $/48$. Finally, all remaining prefix lengths, from $/49$ to $/64$ are left in the third prefix length sorting group.

For each of the $K$ PLS group created, an HTT is built. Thus, the lookup step associated to the prefix length sorting method consists of traversing the $K$ HTTs held in the selected address block bin. 

To summarize, the created PLS groups cover disjoint prefix length ranges by construction. Therefore, the PLS method directly reduces prefix overlaps in each address block bin that increases the performance of HTT. However, within each PLS group, the prefix density variation remains uneven. Hence, a hybrid-trie tree is proposed that exploits the local prefix characteristics to build an efficient data structure.

\section{Hybrid Trie-Tree data structure}\label{sec:trie_tree_hybrid_data_structure}

The hybrid trie-tree proposed in this work is designed to leverage the prefix density variation. This hybrid data structure uses a density-adaptive trie, and a reduced D-Tree leaf when the number of prefixes covered by a density-adaptive trie node is below a fixed threshold value. A description of the two data structures is first presented, then the procedure to build the hybrid trie-tree is formulated, and finally the lookup procedure is introduced.

\subsection{Density-Adaptive Trie} 

The proposed density-adaptive trie is a data structure that is built upon the prefix density distribution. A density-adaptive trie combines a trie data structure with the Selective Node Merge (SNM) method.

While a trie or multi-bit trie creates equi-sized regions whose size is independent of the prefix density distribution, the proposed SNM method adapts the size of the equi-sized regions to the prefix density. Low-density equi-sized regions created with a trie data structure are merged into variable region sizes by the SNM method. Two equi-sized regions are merged if the total number of prefixes after merging is equal to the largest number of prefixes held by the two equi-sized regions, or if it is less than a fixed threshold value. The SNM method merges equi-sized regions both from the low indices to the highest ones and from the high indices to the lowest ones. For both directions, the SNM method first selects the lowest and highest index equi-sized regions, respectively. Second, it evaluates if each selected region can be merged with its next contiguous equi-sized region. The two steps are repeated until the selected region can no longer be merged with its next contiguous equi-sized region. Else, the two previous steps are repeated from the last equi-sized region that was left un-merged. The SNM method has two constraints with respect to the number of merged regions. First, each merged region covers a number of equi-sized regions that is restricted to powers of two, as the space covered by the merged region is described with the prefix notation. Second, the total number of merged regions is bounded by the size of a node header field of the adaptive trie. 
By merging equi-sized regions together, the SNM method reduces the number of regions that are stored in the data structure. As a result, the SNM method improves the memory efficiency of the data structure.

\begin{figure}
\subfloat[First level of a trie data structure with Selective Node Merge]{
  \includegraphics[scale=0.6]{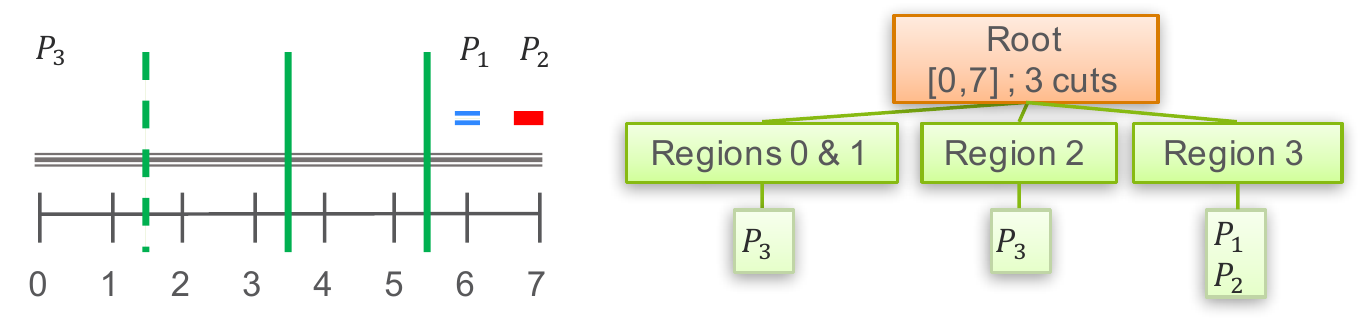}
  \label{fig:with_selective_node_merge}

}
\newline
\subfloat[First level of a trie data structure without Selective Node Merge]{
  \includegraphics[scale=0.6]{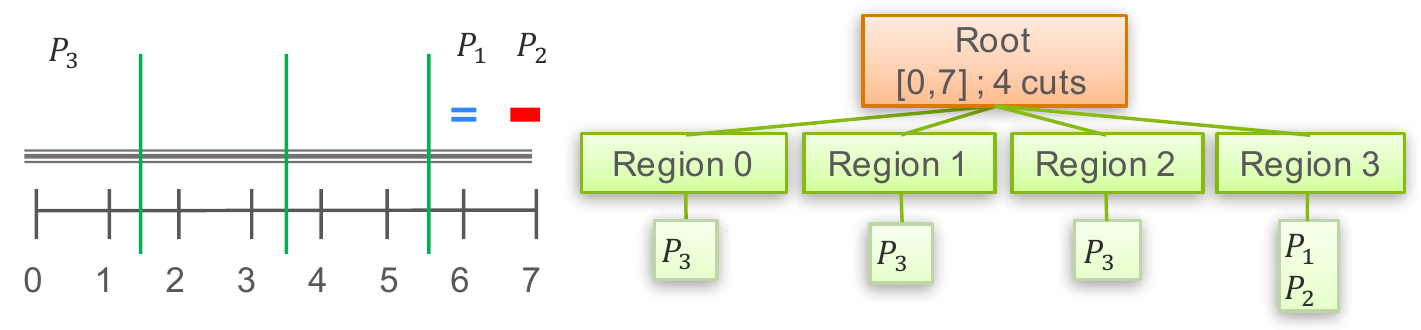}
    \label{fig:wo_selective_node_merge}
}
   \caption{Impact of Selective Node Merge on the replication factor for the first level of a trie data structure.}
   \label{fig:selective_node_merge_example}
\end{figure}

The benefit of the SNM method on the memory efficiency is presented in Fig.~\ref{fig:with_selective_node_merge} for the first level of a multi-bit trie. As a reference, the first level of a multi-bit trie without the SNM method is also presented in Fig.~\ref{fig:wo_selective_node_merge}. In both figures, IP addresses are defined on $3$ bits for the following prefix set $P_{1} = 110/3$, $P_{2} = 111/3 $ and $P_{3} = 0/0$. In both figures, the region is initially partitioned into four equi-sized regions, each corresponding to a different bit combination, called $0$ to $3$. In Fig.~\ref{fig:with_selective_node_merge}, the SNM method merges the two leftmost equi-sized regions $0$ and $1$, separated by a dashed line, as they fulfill the constraints of SNM. In Fig.~\ref{fig:with_selective_node_merge}, not only the SNM method reduces the number of nodes held in memory by 25\% compared to the multi-bit trie presented in Fig.~\ref{fig:wo_selective_node_merge} but also prefix $P_{3}$ is replicated twice, that is a 33\% reduction of the prefix replication factor. As a result, the SNM method increases the memory efficiency of the multi-bit trie data structures.

The regions that are traversed by the SNM method (merged or not) are stored in a SNM field of the adaptive-trie node header. The SNM field is divided into a $LtoH$ and $HtoL$ array. The $LtoH$ and $HtoL$ arrays hold the indices of the regions traversed respectively from low to high index values, and high to low index values. For each region traversed by the SNM method, merged or equi-sized, one index is stored either in the $LtoH$ or the $HtoL$ array. Indeed, as a merged region holds two or more multiple contiguous equi-sized regions, a merged region can be described with the indices of the first and the last equi-sized region it holds. In addition, the SNM method traverses the equi-sized regions contiguously. Therefore, the index of the last equi-sized region held in a merged region can be determined implicitly using the index of the next region traversed by the SNM method. The index value of a non-merged region is sufficient to fully describe it. 

\subsection{Reduced D-Tree leaf}

A reduced D-Tree leaf is built when the number of prefixes held in a region of the density-adaptive trie is below a fixed threshold value  \textit{b}. The proposed leaf is based on a D-tree leaf~\cite{hicuts,scalable_packet_classification} that is extended with the Leaf Size Reduction technique (LSR). 

A D-Tree leaf is a bucket that stores the prefixes and their associated NHI held in a given region. A D-Tree leaf has a memory complexity and time complexity of $O(n)$ for $n$ prefixes stored. A D-Tree leaf is used for the regions at the bottom of the density-adaptive trie because of its higher memory efficiency in those regions. Indeed, we observed that most of the bottom level regions of an density-adaptive trie hold highly unevenly distributed prefixes. Moreover, a D-Tree leaf has better memory efficiency with highly unevenly distributed prefixes over a density-adaptive trie. Whereas a density-adaptive trie can create prefix replication, which reduces the memory efficiency, no prefix replication is created with a D-Tree leaf. However, the D-Tree leaf comes at the cost of higher time complexity compared to a density-adaptive trie.

As a consequence, the LSR technique is introduced to reduce the time complexity of a D-Tree leaf by reducing the amount of information stored in a D-Tree leaf. In fact, a D-Tree leaf stores entirely each prefix even the bits that have already been matched by the density-adaptive trie before reaching the leaf. On the other hand, the LSR technique stores in the reduced D-Tree leaf only the prefix bits that are left unmatched. To specify the number of bits that are left unmatched, a new LSR leaf header field is added, coded on $6$ bits. The LSR technique reduces the amount of information that is stored in each reduced D-Tree leaf. As a result, not only does the reduced D-Tree leaf requires fewer memory accesses but it also has a better memory efficiency over a D-Tree leaf.

\subsection{HTT build procedure}

The hybrid trie-tree build procedure is presented in Algorithm~\ref{alg:hybrid_data_structure_build_proc}, starting with the root region that holds all the prefixes (line $1$). If the number of prefixes stored in the root region is below a fixed threshold \textit{b}, then a reduced D-Tree leaf is built (line $2 - 3$). Else, the algorithm iteratively partitions this region into equi-sized regions (lines $4 - 10$). The SNM method is then applied on the equi-sized regions (line $11$). Next, for each region, if the number of prefixes is below the threshold value (line $12$), a reduced D-Tree leaf is built (line $13$), else a density-adaptive trie node is built (line $14 - 16$) and the region is again partitioned.
\begin{algorithm}
\caption{Hybrid Trie-Tree build procedure}
\label{alg:hybrid_data_structure_build_proc}
\begin{algorithmic}[1]
\renewcommand{\algorithmicrequire}{\textbf{Input: }}
\renewcommand{\algorithmicensure}{\textbf{Output: }}
\Require Prefix Table, stack Q
\Ensure  Hybrid Trie-Tree
\State{Create a root region covering all prefixes \;}
\If{the number of prefixes held in that region is below the threshold value \textit{b}}
	\State Create a reduced D-Tree leaf for those prefixes \;
\Else
	\State Push the root region onto Q \;
\EndIf
\While{Q is not empty}
\parState{Remove the top node in Q and use it as the reference region}
\parState{Compute the number of partitions in the reference region}
\parState{Partition the reference region according to the previous step}
\parState{Apply the SNM method on the partitioned reference regions}
\For{each partitioned reference region}
	\If{it holds a number of prefixes that is below the threshold value}
		\State{Create a reduced D-Tree leaf for those prefixes}
	\Else
		\parState{Build an adaptive-density trie node for those prefixes}
		\State Push this region onto Q
\EndIf
\EndFor
\EndWhile
\State \Return the Hybrid Trie-Tree \;
\end{algorithmic}
\end{algorithm}

The number of partitions in a region (line $9$ of Algorithm~\ref{alg:hybrid_data_structure_build_proc}) is computed by a greedy heuristic proposed in~\cite{hicuts}. The heuristic uses the prefix distribution to adapt the number of partitions, as expressed in Algorithm~\ref{alg:bi_heuristic}. An objective function, the \textit{Space Measurement (Sm)} is evaluated at each iteration (lines $4$ and $5$) and compared to a threshold value, the \textit{Space Measurement Factor (Smpf)} evaluated in the first step (line $1$). The number of partitions increases by a factor of two at each iteration (line $3$), until the value of the objective function \textit{Sm} (line $4$) becomes greater than the threshold value (line $5$). The objective function estimates the memory usage efficiency with the prefix replication factor by summing the number of prefixes held in each $j$ equi-sized region created $\sum_{j=0}^{N_{p}} Num_{Prefixes}(equi-sized region_{j} )$ (line $4$). The prefix replication factor is impacted by the prefix distribution. If prefixes are evenly distributed, the replication factor remains very low until the equi-sized regions become smaller than the average prefix size. Then, the prefix replication factor increases exponentially. Thus, to avoid over-partitioning a region if the replication factor remains low for many iterations, the number of partitions $N_{p}$ and the result of the previous iterations $Sm(N_{p-1})$ are used as a penalty term that is added to the objective function (line $4$). On the other hand, if prefixes are unevenly distributed, the prefix replication factor increases linearly until the largest prefixes in the region partitioned become slightly smaller compared to an equi-sized region. Passed this point, an exponential growth of the replication factor is observed. The heuristic creates fine-grained partition size in a dense region, and coarse-grained partition size in a sparse region. 

\begin{algorithm}
\caption{Heuristic used to compute the number of partitions in a region}
\label{alg:bi_heuristic}
\begin{algorithmic}[1]
\renewcommand{\algorithmicrequire}{\textbf{Input: }}
 \renewcommand{\algorithmicensure}{\textbf{Output: }}
 \Require Region to be cut
 \Ensure  Number of partitions ($N_{p}$)

\label{alg:bi_obj_greedy}
\State $N_{p} = 1; Smpf = Num_{Prefixes} \cdot 8;  Sm(N_{p}) = 0 ;$
\Do  
\State $N_{p} = N_{p} \cdot 2; $
\parState{$ Sm(N_{p})= \sum_{j=0}^{N_{p}} Num_{Prefixes}(equi-sized region_{j})  + N_{p} + Sm(N_{p-1}) ; $}
\doWhile $Sm(N_{p}) \leq Smpf   $ \\
\Return $N_{p}$
\end{algorithmic}
\end{algorithm}

The number of partitions in a region is a power of two. Thus, the base-2 logarithm of the number of partitions represents the number of bits from the IP address used to select the equi-sized region covering this IP address.

\subsection{HTT lookup procedure}\label{sec:lookup_fixed}

The hybrid trie-tree lookup algorithm starts with a traversal of the density-adaptive trie until a reduced D-Tree leaf is reached. Next, the reduced D-Tree leaf is traversed to identify the matching prefix and its NHI. 

The traversal of the density-adaptive trie consists in computing the memory address of the child node that matches the destination IP address, calculated with Algorithm~\ref{alg:children_node_address}. This algorithm uses as input parameters the memory base address, the destination-IP bit-sequence, the $LtoH$ and the $HtoL$ arrays that are extracted from the node header. The SNM method can merge multiple equi-sized nodes into a single node in memory, and thus the destination-IP bit-sequences cannot be used directly as an index to the child node. Therefore, Algorithm~\ref{alg:children_node_address} computes for each destination-IP bit-sequence the number of equi-sized nodes that are skipped in memory based on the characteristics of the merged regions described in the $LtoH$ and the $HtoL$ arrays. The value of the destination-IP bit-sequence can point to a region that is either included 1) in a merged region described in the $LtoH$ array (line $1$), or 2) in a merged region described in the $HtoL$ array (line $4$), or 3) in a equi-sized region that has not been traversed by the SNM method (line $7$). 

The following notation is introduced: $L$ represents the size of the $HtoL$ and $LtoH$ arrays, $LtoH [i]$ and $HtoL [i]$ are respectively the $i-th$ entry of the $LtoH$ and the $HtoL$ arrays. In the first case, each entry of the $LtoH$ array is traversed to find the closest $LtoH [i]$ that is less than or equal to the destination-IP bit-sequence (line $1$). The index of the matched child node is equal to $index_{LtoH}$ (line $2$), where $index_{LtoH}$ is the index of the $LtoH$ array that fulfills this condition. In the second case, each entry of the $HtoL$ array is similarly traversed to find the closest $HtoL [i]$ that is greater than or equal to the destination-IP bit-sequence (line $4$). The $index_{HtoL}$ in the $LtoH$ array that fulfills this condition is combined with the characteristics of the $LtoH$ and $HtoL$ arrays to compute the index of the selected child node (line $5$). In the third case, the algorithm evaluates only the number of equi-sized nodes that are skipped in memory based on the characteristics of the $LtoH$ array and the destination IP address bit sequence of the matched child node (line $7$). Finally, the index that is added to the base address points in memory to the matched child node.

\begin{algorithm}
\caption{Memory address of the matched child node using SNM method}
\label{alg:children_node_address}
\begin{algorithmic}[1]
\renewcommand{\algorithmicrequire}{\textbf{Input: }}
 \renewcommand{\algorithmicensure}{\textbf{Output: }}
 \Require Children base address, destination IP address bit sequence, $LtoH$ and $HtoL$ arrays
  \Ensure  Child node address

\Comment{Index included in a region using SNM method}
\If{destination IP address bit sequence  $\leq  LtoH [L-1]$}
	\Comment{In LtoH array}
	\State{Index = $Index_{LtoH}$}
\Else{
	\If{destination IP address bit sequence $\geq  HtoL[0] $}
		\Comment{In HtoL array}
		\parState{Index = $index_{HtoL} + HtoL [0] - LtoH [L-1]+ L-1$}
 	\Else{
		\Comment{destination IP address bit sequence included in an equi-sized region that has not been traversed by the SNM method}
		\parState{Index = destination IP address bit sequence $- LtoH [L-1]+ L-1$}
		\EndIf}
		}
 	\EndIf
	 \\
 \Return{Child~node~address = base~address + Index}
\end{algorithmic}
\end{algorithm}

\begin{figure}[htbp]
	\centering
	\subfloat[Index in memory of the regions before and after the SNM method is applied]{{\includegraphics[width=0.65\linewidth]{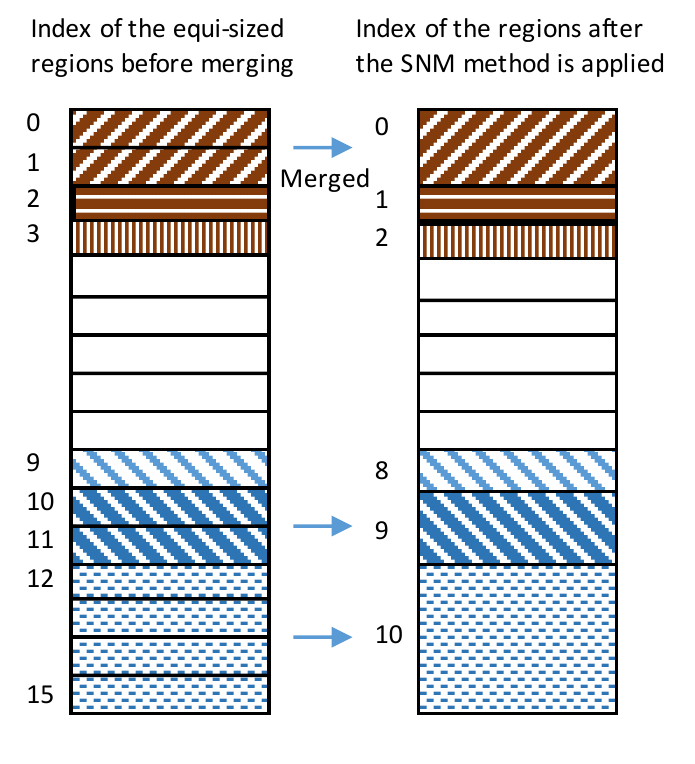} }}%
    \qquad
    \subfloat[The SNM field associated to the regions traversed by the SNM method]{{\includegraphics[width=0.22\linewidth]{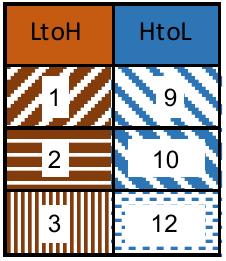} }}%
    \caption{SNM method applied to a region that holds 11 nodes after merging, and its associated SNM field}%
    \label{fig:example_SNM}%
\end{figure}

Algorithm~\ref{alg:children_node_address} is illustrated with Figures~\ref{fig:example_SNM} in which $L = 3$ and the destination IP address bit sequence is arbitrarily set to $10$. Based on Fig.~\ref{fig:example_SNM}, the destination IP address bit sequence $10$ matches the equi-sized region with the index $10$ before the SNM method is applied. However, after the SNM method is applied, the destination IP address bit sequence matches a merged node with the index $9$. Based on the SNM header, the destination IP address bit sequence $10$ is greater than both $LtoH [L-1]= 3$ and $HtoL [0]= 9$. Thus, we must identify the number of equi-sized nodes that are skipped in memory with the $LtoH$ and $HtoL$ arrays. Because $LtoH [L-1] = 3$, two equi-sized nodes have been merged. As one node is skipped in memory, any child index greater than $3$ is stored at offset $index - 1$ in memory. Moreover, the destination IP address bit sequence is greater than $HtoL [0]= 9$. However, $HtoL [1] = 10$, meaning that indices $9$ and $10$ are not merged, and no entry is skipped in memory for the first two regions held in the $HtoL$ array. As a consequence, only one node is skipped in memory, and thus the child node index is $10 - 1 = 9$.

\begin{algorithm}
\caption{Lookup in the Reduced D-tree leaf}
\label{alg:lookup_reduced_leaf}
\begin{algorithmic}[1]
\renewcommand{\algorithmicrequire}{\textbf{Input: }}
 \renewcommand{\algorithmicensure}{\textbf{Output: }}
 \Require Reduced D-Tree leaf, destination IP Address
 \Ensure  LPM and its NHI
\State {Parse the leaf header}
\State {Read the prefixes held in the leaf}
\For {each prefix held in the leaf}
\parState{Match the destination IP address against the selected prefix}
\If{Positive Match}
\State{Record the prefix length of the matched prefix}
\EndIf 
\EndFor 
\State{Identify the longest prefix match amongst all positive matches}\\
\Return {the longest prefix match and its NHI }
\end{algorithmic}
\end{algorithm}

The density-adaptive trie is traversed until a reduced D-Tree leaf is reached. The lookup procedure of a reduced D-Tree leaf is presented in Algorithm~\ref{alg:lookup_reduced_leaf}. The leaf header is first parsed, and then prefixes are read (lines $1$ to $2$). Next, all prefixes are matched against the destination IP address, and their prefix length is recorded if matches are positive (lines $3$ to $6$). When all the prefixes are matched, only the longest prefix match is returned with its NHI (lines $7-8$).

\section{Performance Measurement Methodology}\label{sec:method}

This section describes the methodology used to evaluate SHIP performance using both real and synthetic prefix tables. Eleven real prefix tables were extracted using the RIS remote route collectors~\cite{ris_raw_data}, and each one holds approximately $25$ k prefixes. Each scenario, noted $rrc$ followed by a two-digit number, characterizes the location in the network of the remote route collector used. For prefix tables holding up to $580$ k entries, synthetic prefixes were generated with a method that uses IPv4 prefixes to generate IPv6 prefixes, in a one-to-one-mapping~\cite{non_random_generation}. The IPv4 prefixes used were also extracted from~\cite{ris_raw_data}. Using the IPv6 prefix table holding $580$ k prefixes, four smaller prefix tables were created, with a similar prefix length distribution, holding respectively $290$ k, $116$ k, $58$ k and $29$ k prefixes. 

The performance of SHIP was evaluated using two metrics: the number of memory accesses to traverse its data structure and its memory consumption. For the two metrics, the performance is reported separately for the hash table used by the address block binning method, and the HTTs built by the prefix length sorting method. SHIP performance is characterized using $1$ to $6$ groups for two-level prefix grouping, and as a reference the performance of a single HTT without grouping is also presented.The number of groups is limited to six, as we have observed with simulations that increasing it further does not improve the performance.

For the evaluation of the number of memory accesses, it is assumed that the selected hybrid trie-trees within an address block bin are traversed in parallel, using dedicated traversal engines. Therefore, the reported number of memory accesses is the largest number of memory accesses of all the hybrid trie-trees amongst all address block bins. It is also assumed that the memory bus width is equal to a node size, in order to transfer one node per memory clock cycle. 

The memory consumption is evaluated as the sum of all nodes held in the hybrid trie-tree for the prefix length sorting method, and of the size of the perfect hash table used for the address block binning method. In order to evaluate the data structure overhead, this metric is given in bytes per byte of prefix. This metric is evaluated as the size of the data structure divided by the size of the prefixes held in the prefix table.
The format and size of a non-terminal node and a leaf header used in a hybrid trie-tree are detailed respectively in Table~\ref{tab:node_format} and in Table~\ref{tab:SHIP_leaf_header}. The node type field, coded with $1$ bit, specifies whether the node is a leaf or a non-terminal node. The following fields are used only for non-terminal nodes. Up to $10$ bits can be matched at each node, corresponding to a node header field coded with $4$ bits.  The fourth field is used for SNM, to store the index value of the traversed regions. Each index is restricted to $10$ bits, while the $HtoL$ and $LtoH$ arrays each store up to $5$ indices. The third field, coded in $16$ bits, stores the base address of the first child node associated with its parent's node. 

 \begin{table}[htb]
\renewcommand{\arraystretch}{1.3}
\caption{ Non-terminal node header field sizes in bits}
 \label{tab:node_format}
 \centering
      \begin{tabular}{|c|c|c|}
   \hline
\bf Header Field  &  \bf   Size \\
      \hline
  Node type &     $1$\\
\hline 
Number of cuts  &  $4$  \\
\hline
Pointer to child node  & $16$ \\
\hline 
Size of selective node merge array    &  $5 \cdot 10  + 5 \cdot 10$ \\
\hline 
    \end{tabular}
\end{table}

The leaf node format is presented in Table~\ref{tab:SHIP_leaf_header}. A leaf can be split over multiple nodes to store all its prefixes. Therefore, two bits are used in the leaf header to specify whether the current leaf node is a terminal leaf or not. The next field gives the number of prefixes stored in the leaf. It is coded with $4$ bits because in this work, the largest number of prefixes held in a leaf is set to $12$ for each hybrid trie-tree. The LSR field stores the number of bits that need to be matched, using 6 bits. If a leaf is split over multiple nodes, a pointer coded with $16$ bits points at the remaining nodes that are part of the leaf. Inside a leaf, prefixes are stored alongside their prefix length and with their NHI. The prefix length is coded with the number of bits specified by the LSR field while the NHI is coded with $8$ bits.

\begin{table}[htb]
\renewcommand{\arraystretch}{1.3}
\caption{Leaf header field sizes in bits}
 \label{tab:SHIP_leaf_header}
 \centering
      \begin{tabular}{|c|c|c|} 
   \hline
\bf Header Field    & Size    \\
      \hline
  Node type & $2$  \\
\hline
Number of prefixes stored & $4$ \\
\hline 
LSR field &  $6$   \\
\hline
Pointer to remaining leaf entries &  $16$  \\
\hline 
Prefix and NHI  & Value specified in the LSR field + $8$  \\
\hline 
    \end{tabular}
\end{table}

\section{Results}\label{sec:results}

SHIP performance is first evaluated using real prefixes, and then with synthetic prefixes, for both the number of memory accesses and the memory consumption.

\subsection{Real Prefixes}
The performance analysis is first made for the perfect hash table used by the address block binning method. In Table~\ref{tab:result_bin23_real},
the memory consumption and the number of memory accesses for the hash table are shown. The ABB method uses between $19$ kB and $24$ kB, that is between $0.7$ and $0.9$ bytes per prefix byte for the real prefix tables evaluated. The memory consumption is similar across all the scenarios tested as prefixes share most of the $23$ MSBs. On the other hand, the number of memory accesses is by construction independent of the number of prefixes used, and constant to $2$. 

\begin{table}[htbp]
\renewcommand{\arraystretch}{1.3}
\caption{Memory consumption of the address block binning method for real prefix tables}
 \label{tab:result_bin23_real}
 \centering
      \begin{tabular}{|l|c|c|}
   \hline
\bf Scenario    & \bf Hashing Table size (kB)  & \bf Memory Accesses   \\
\hline
$rrc00$ & 20  &	2 \\
\hline
$rrc01$ & 19  &	2 \\
\hline
$rrc04$ & 24  &	2 \\
\hline
$rrc05$ & 19  &	2 \\
\hline
$rrc06$ & 19  &	2 \\
\hline
$rrc07$ & 20  &	2 \\
\hline
$rrc10$ & 21  &	2 \\
\hline
$rrc11$ & 20  &	2 \\
\hline
$rrc12$ & 20  &	2 \\
\hline
$rrc13$ & 22  &	2 \\
\hline
$rrc14$ & 21  &	2 \\
\hline 
    \end{tabular}
\end{table}

\begin{figure}
\centering
\subfloat[Memory Consumption]{\includegraphics[width=3.5in]{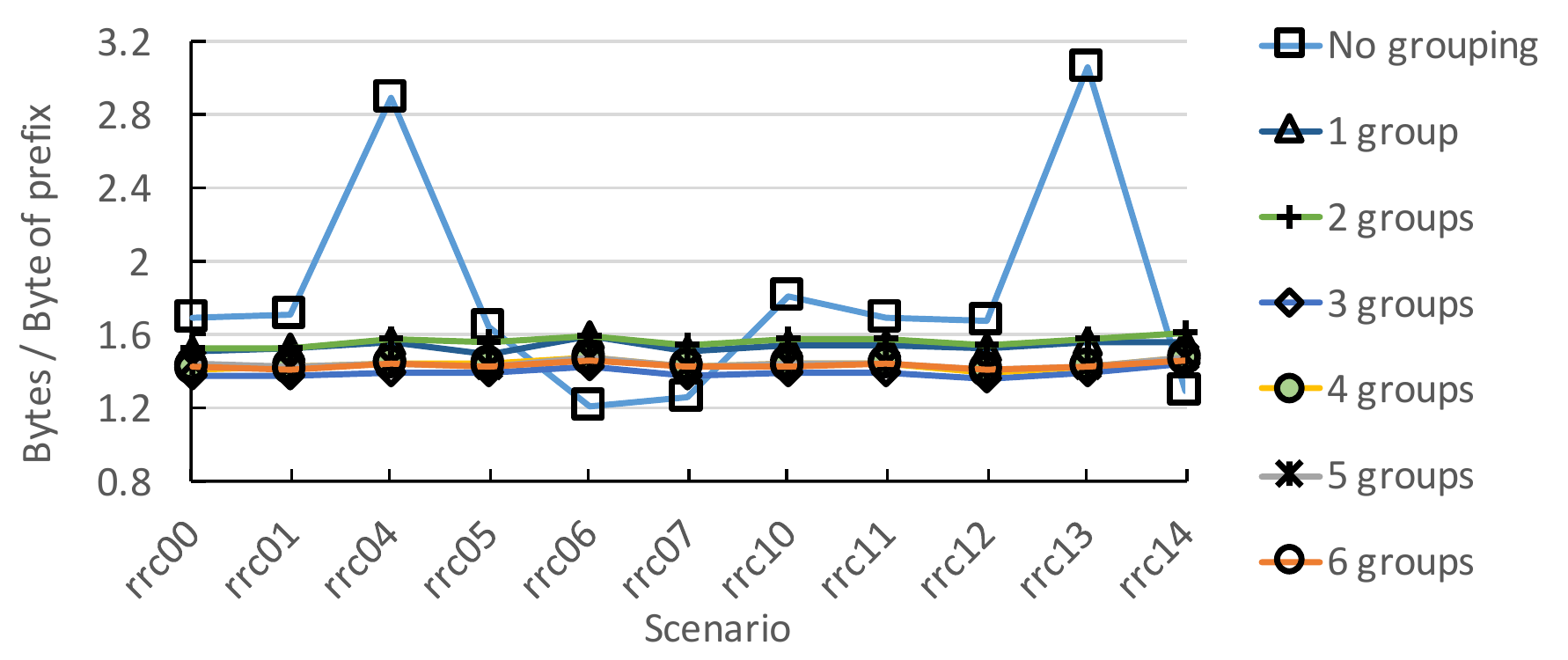}
\label{mem_cons_real_prefix_table}}
\hfil
\subfloat[Memory Accesses]{\includegraphics[width=3.5in]{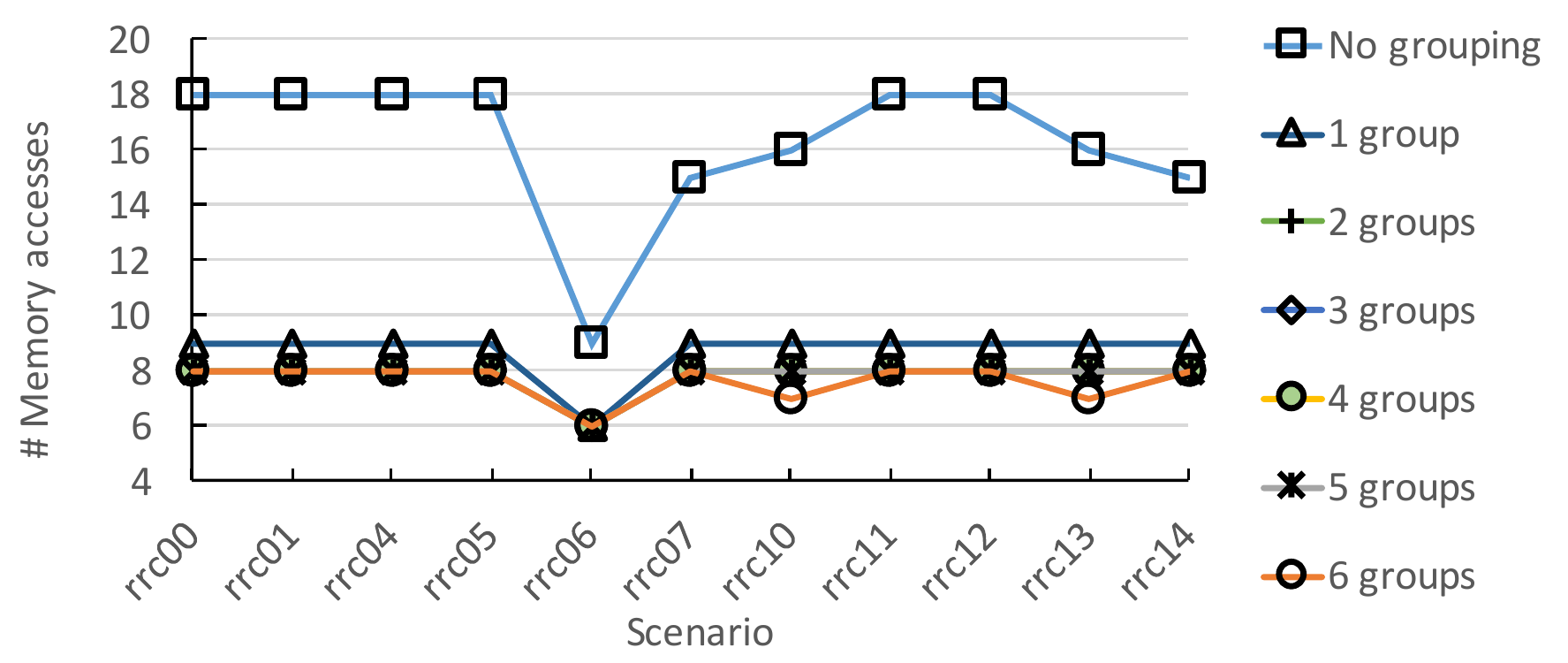}
\label{mem_acc_real_prefix_table}}
\caption{Real prefix tables: impact of the number of groups on the memory consumption (a) and the number of memory accesses (b) of the HTTs.}
\label{fig:result_memory_consumption}
\end{figure}

In Figures~\ref{mem_cons_real_prefix_table} and~\ref{mem_acc_real_prefix_table}, the performance of the HTTs is evaluated respectively on the memory consumption and the number of memory accesses. In both figures, $1$ to $6$ groups are used for two-level prefix grouping. As a reference, the performance of the HTT without grouping is also presented.

In Fig.~\ref{mem_cons_real_prefix_table}, the memory consumption of the HTTs ranges from $1.36$ to $1.60$ bytes per prefix byte for all scenarios, while it ranges between $1.22$ up to $3.15$ bytes per byte of prefix for a single HTT. Thus, using two-level prefix grouping, the overhead of the HTTs ranges from $0.36$ to $0.6$ byte per byte of prefix. However, a single HTT leads to an overhead of $0.85$ on average, and up to $3.15$ bytes per byte of prefix for scenario $rrc13$. Thus, two-level grouping reduces the memory consumption and smooths its variability, but it also reduces the hybrid trie-tree overhead. 
 
Fig.~\ref{mem_cons_real_prefix_table} shows that increasing the number $K$ of groups up to three reduces the memory consumption. However, using more groups does not improve the memory consumption, and even worsens it. Indeed, it was observed experimentally that when increasing the value of $K$ most groups hold very few prefixes, leading to a hybrid trie-tree holding a single leaf with part of the allocated node memory left unused. Thus, using too many groups increases memory consumption.

It can be observed in Fig.~\ref{mem_acc_real_prefix_table} that the number of memory accesses to traverse the HTTs ranges from $6$ to $9$ with two-level prefix grouping, whereas it varies between $9$ and $18$ with a single HTT. So, two-level prefix grouping smooths the number of memory accesses variability, but it also reduces on average the number of memory accesses approximatively by a factor 2.
 
However, increasing the number $K$ of groups used by two-level prefix grouping, from $1$ to $6$, yields little gain on the number of memory accesses, as seen in Fig.~\ref{mem_acc_real_prefix_table}. Indeed, for most scenarios, one memory access is saved, and up to two memory accesses are saved in two scenarios, by increasing the number $K$ of groups from $1$ to $6$. Indeed, for each scenario, the performance is limited by a prefix length that cannot be divided in smaller sets by increasing the number of groups. Still, using two or more groups, in the worst case, $8$ memory accesses are required for all scenarios. The performance is similar across all scenarios evaluated, as few variations exist between the prefix groups created using two-level grouping for those scenarios.


\subsection{Synthetic Prefixes}

The complexity of the perfect hash table used for the address block binning method is presented in Table~\ref{tab:result_bin23} with synthetic prefix tables. It requires on average $2.7$ bytes per prefix byte for the $5$ scenarios tested, holding from $29$ k up to $580$ k prefixes. The perfect hash table used shows linear memory consumption scaling. For the number of memory accesses, its value is independent of the prefix table, and is equal to $2$. 

\begin{table}[htbp]
\renewcommand{\arraystretch}{1.3}
\caption{Cost of binning on the first 23 bits for synthetic prefix tables}
 \label{tab:result_bin23}
 \centering
      \begin{tabular}{|l|c|c|}
   \hline
\bf Prefix Table Size    & \bf Hashing Table size (kB)  & \bf Memory Accesses   \\
\hline
580 k &  1282 & 2 \\
\hline 
290 k & 642 &	2 \\
\hline 
110 k & 322 &	2 \\
\hline 
50 k & 162 &	2 \\
\hline 
29 k & 82 &	2 \\
\hline 
    \end{tabular}
\end{table}

\begin{figure}[htbp]
\centering
\subfloat[Memory accesses]{\includegraphics[width=3.5in]{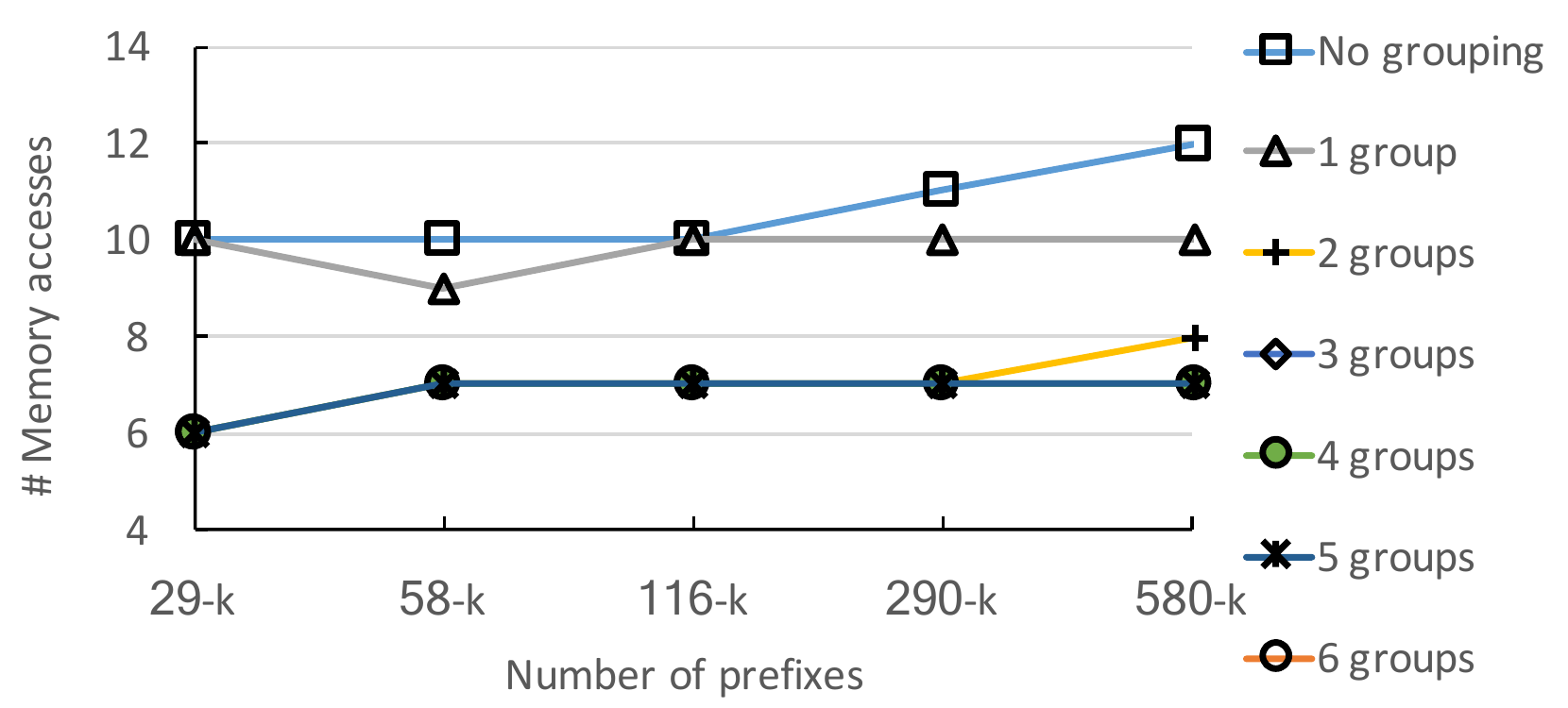}%
\label{mem_acc_synth_prefix_table}}
\hfil
\subfloat[Memory consumption]{\includegraphics[width=3.5in]{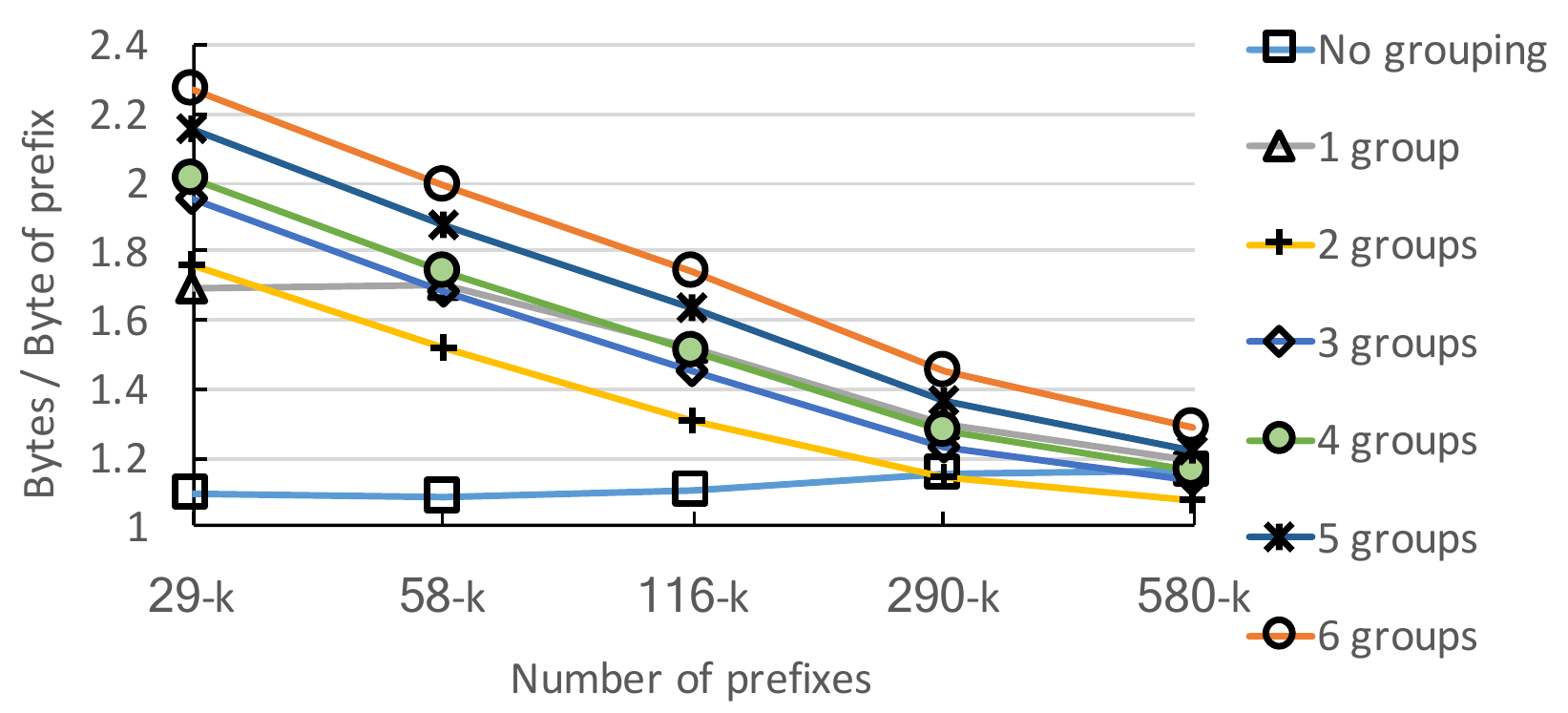}
\label{mem_cons_synth_prefix_table}}
\hfil
\subfloat[Memory consumption scaling]{\includegraphics[width=3.5in]{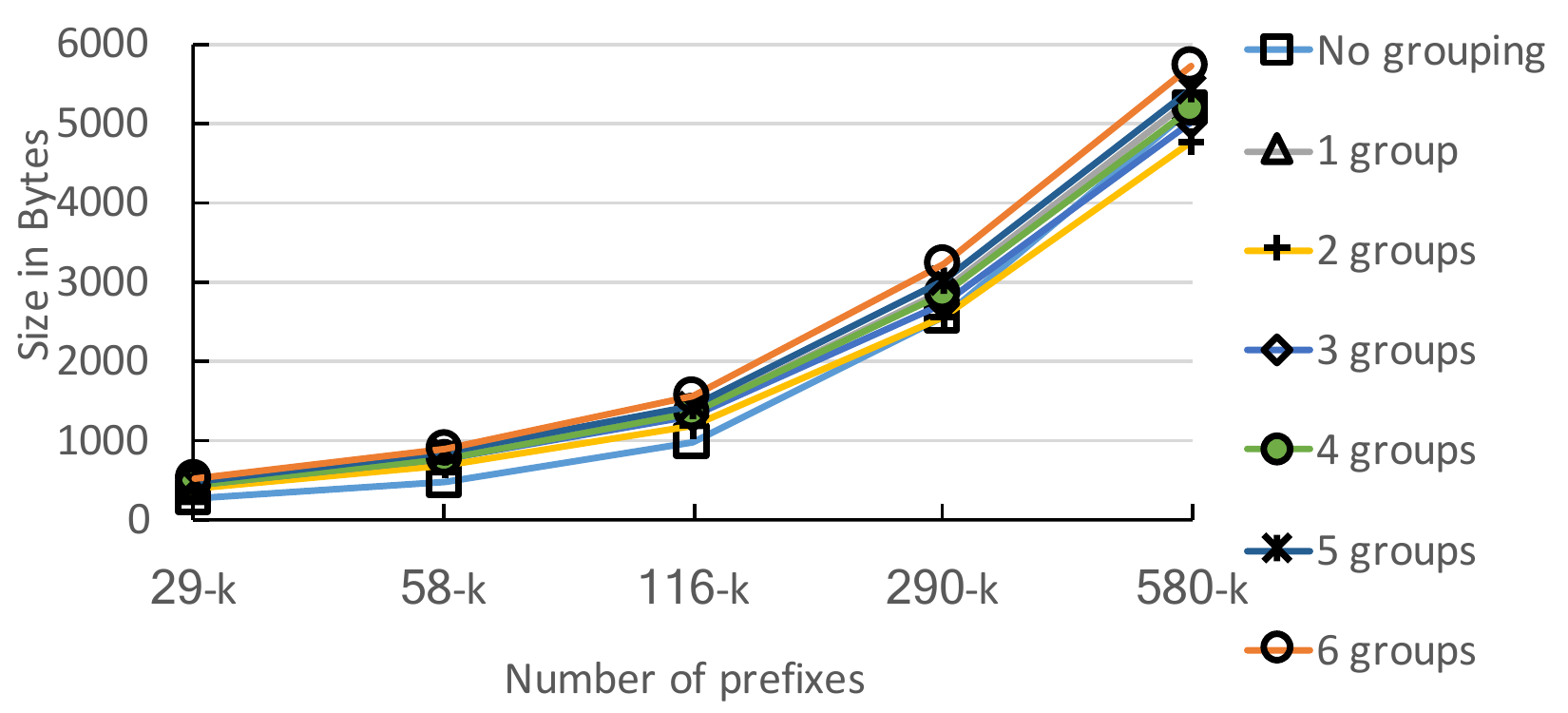}
\label{fig:mem_cons_synth_prefix_table_scaling}}

\caption{Synthetic prefix tables: impact of the number of groups on the number of memory accesses (a), the memory consumption (b) and scaling (c) of the HTTs.}
\label{fig:result_memory_accesses}
\end{figure}

The performance of the HTTs with synthetic prefixes is evaluated for the number of memory accesses, the memory consumption, and the memory consumption scaling, respectively in Fig.~\ref{mem_acc_synth_prefix_table},~\ref{mem_cons_synth_prefix_table}, and~\ref{fig:mem_cons_synth_prefix_table_scaling}. For each of the three figures, $1$ to $6$ groups are used for two-level prefix grouping. The performance of the HTT without grouping is also presented in the three figures, and is used as a reference.

Two behaviors can be observed for the memory consumption in Fig.~\ref{mem_cons_synth_prefix_table}. 
First, for prefix tables with $290$ k prefixes and more, it can be seen that two-level prefix grouping used with $2$ groups slightly decreases the memory consumption over a single HTT. Using this method with two groups, the HTTs consumes between $1.18$ and $1.09$ byte per byte of prefix, whereas the memory consumption for a single HTT lies between $1.18$ and $1.20$ byte per byte of prefix. However, increasing the number of groups to more than two does not improve memory efficiency, as it was observed that most prefix length sorting groups hold very few prefixes, leading to hybrid trie-tree holding a single leaf, with part of the allocated node memory that is left unused. Even though the memory consumption reduction brought by two-level prefix grouping over a single HTT is small for large synthetic prefix tables, it will be shown in this paper that the memory consumption remains lower when compared to other solutions. Moreover, it will be demonstrated that two-level prefix grouping reduces the number of memory accesses to traverse the HTT with the worst case performance over a single HTT, for all synthetic prefix table sizes. Second, for smaller prefix tables with up to $116$ k prefixes, a lower memory consumption is achieved using only a single HTT for two reasons. First, the synthetic prefixes used have fewer overlaps and are more distributed than real prefixes for small to medium size prefix tables, making two-level prefix grouping less advantageous in terms of memory consumption. Indeed, a larger number $M$ of address block bins has been observed compared to real prefix tables with respect to the number of prefixes held the prefix tables, for small and medium prefix tables. Thus, on average, each bin holds fewer prefixes compared to real prefix tables. As a consequence, we observe that the average and maximum number of prefixes held in each PLS group is smaller for prefix tables holding up to $116$ k prefixes. It then leads to hybrid trie-trees where the allocated leaf memory is less utilized, achieving lower memory efficiency and lower memory consumption.

In order to observe the memory consumption scaling of the HTTs, Fig.~\ref{fig:mem_cons_synth_prefix_table_scaling} shows the total size of the HTTs using synthetic prefix tables, two-level prefix grouping, and a number $K$ of groups that ranges from $1$ to $6$. The memory consumption of the HTTs with and without two-level prefix grouping grows exponentially for prefix tables larger than $116$ k. However, because the abscissa uses a logarithmic scale, the memory consumption scaling of the proposed HTT is linear with and without two-level prefix grouping. In addition, the memory consumption of the HTTs is $4,753$ kB for the largest scenario with $580$ k prefixes, two-level prefix grouping, and $K = 2$.

Next, we analyze in Fig.~\ref{mem_acc_synth_prefix_table} the number of memory accesses required to traverse the HTT leading to the worst-case performance, for synthetic prefix tables, with two-level prefix grouping using $1$ to $6$ groups. It can be observed that two-level prefix grouping reduces the number of memory accesses over a single HTT, for all the number of groups and all prefix table sizes. The impact of two-level prefix grouping is more pronounced when using two groups or more, as the number of memory accesses is reduced by 40\% over a single HTT. Using more than $3$ groups does not further reduce the number of memory accesses as the group leading to the worst-case scenario cannot be reduced in size by increasing the number of groups. Finally, it can be observed in Fig.~\ref{mem_acc_synth_prefix_table} that the increase in the number of memory accesses for a search is at most logarithmic with the number of prefixes, since each curve is approximately linear and the x-axis is logarithmic.

The performance analysis presented for synthetic prefixes has shown that two-level prefix grouping improves the performance over a single HTT for the two metrics evaluated. Although the performance improvement of the memory consumption is limited to large prefix tables, using few groups, the number of memory accesses is reduced for all prefix table sizes and for all numbers of groups. In addition, it has been observed experimentally that the HTTs used with two-level prefix grouping have a linear memory consumption scaling, and a logarithmic scaling for the number of memory accesses. The hash table used in the address block binning method has shown to offer a linear memory consumption scaling and a fixed number of memory accesses. Thus, SHIP has a linear memory consumption scaling, and a logarithm scaling for the number of memory accesses.

\section{Discussion}\label{sec:disscusion}

This section first demonstrates that SHIP is optimized for a fast hardware implementation. Then, the performance of SHIP is compared with previously reported results.

\subsection{SHIP hardware implementability}

We demonstrate that SHIP is optimized for a fast hardware implementation, as it complies with the following two properties; 1) pipeline-able and parallelizable processing to maximize the number of packets forwarded per second, 2) use of a data structure that can fit within on-chip memory to minimize the total memory latency.

A data structure traversal can be pipelined if it can be decomposed into a fixed number of stages, and for each stage both the information read from memory and the processing are fixed. First, the HTT traversal can be decomposed into a pipeline, where each pipeline stage is associated to a HTT level. Indeed, the next node to be traversed in the HTT depends only on the current node selected and the value of the packet header. Second, for each pipeline stage of the HTT both the information read from memory and the processing are fixed. Indeed, the information is stored in memory using a fixed node size for both the adaptive-density trie and the reduced D-Tree leaf. In addition, the processing of a node is constant for each data structure and depends only on its type, as presented in Section~\ref{sec:lookup_fixed}. As a result, the HTT traversal is pipeline-able. Moreover, the HTTs within the $K$ PLS groups are independent, thus their traversal is by nature parallelizable. As a consequence, by combining a parallel traversal of the HTTs with a pipelined traversal of each HTT, property $1$ is fulfilled. The hash table data structure used for the address block binning technique has been implemented in hardware in previous work~\cite{mem_eff}, and thus it already complies with property $1$.

For the second property, SHIP uses $5.9$ MB of memory for $580$ k prefixes, with two-level prefix grouping, and $K = 2$ . Therefore, SHIP data structure can fit within on-chip memory of the current generation of FPGAs and ASICs~\cite{altera,xilinx,pisa}. Hence, SHIP fulfills property $2$. As both the hash table used by two-level prefix grouping, and the hybrid trie-tree comply with properties $1$ and $2$ required for a fast hardware implementation, SHIP is optimized for a fast hardware implementation. 

\subsection{Comparison with previously reported results}

\begin{table*}[htbp]
\renewcommand{\arraystretch}{1.3}  
\caption{Comparison Results}
 \label{tab:result_comparison}
 \centering
      \begin{tabular}{|l|c|c|c|c|}
   \hline
\bf Method    & \bf Memory Consumption  & \bf Latency (ns)   & \multicolumn{2}{|c|}{\bf Complexity}  \\
		& \bf (in bytes per prefix) & & Memory Consumption & Memory latency  \\
		\hline
Tree-based \cite{scalable_ipv6_vk} & $19.0$ & 90 &  $O(N)$ & $O(log_2 (N) ) \leq Latency \leq 2 \cdot O(log_3 (N) )$\\
 
\hline
CLIPS  \cite{mem_eff} & $27.6$ & N/A & N/A & N/A \\ 
\hline 
FlashTrie \cite{flashtrie_journal} &  $124.2$ & 80 & N/A & N/A \\
\hline 
FlashLook \cite{flashlook} 	& $1010.0$ & 90 &   N/A   & N/A \\
 \hline 
  
\bf SHIP  & $10.64$ & 31  & $O(N)$  & $O(log(N))$ \\ 

\hline
\end{tabular}
\end{table*}

Table~\ref{tab:result_comparison} compares the performance of SHIP and previous work in terms of memory consumption and worst case memory latency. If available, the time and space complexity are also shown. In order to use a common metric between all reported results, the memory consumption is expressed in bytes per prefix, obtained by dividing the size of the data structure by the number of prefixes used. The memory latency is based on the worst-case number of memory accesses to traverse a data structure.
For the following comparison, it is assumed that on-chip SRAM memory running at 322 MHz~\cite{scalable_ipv6_vk} is used, and off-chip DDR3-1600 memory running at 200 MHz is used.

Using both synthetic and real benchmarks, SHIP requires in the worst case $10$ memory accesses, and consumes $5.9$ MB of memory for the largest prefix table, with $2$ groups for two-level prefix grouping. Hence, the memory latency to complete a lookup with on-chip memory is equal to $10 \cdot 3.1$ = 31 ns.

FlashTrie has a high memory consumption, as reported in Table~\ref{tab:result_comparison}. The results presented were reevaluated using the node size equation presented in~\cite{flashtrie_conf} due to incoherence with equations shown in~\cite{flashtrie_journal}. This algorithm leads to a memory consumption per prefix that is around $11 \times$ higher than the SHIP method, as multiple copies of the data structure have to be spread over DDR3 DRAM banks. In terms of latency, in the worst case, two on-chip memory accesses are required, followed by three DDR3 memory bursts. However, DRAM memory access comes at a high cost in terms of latency for the FlashTrie method. First, independently of the algorithm, a delay is incurred to send the address off-chip to be read by the DDR3 memory controller. Second, the latency to complete a burst access for a given bank, added to the maximum number of bank-activate commands that can be issued in a given period of time, limits the memory latency to $80$ ns and reduces the maximum lookup frequency to $84$ MHz. Thus, FlashTrie memory latency is $2.5 \times$ higher than SHIP. 

The FlashLook~\cite{flashlook} architecture uses multiple copies of data structures in order to sustain a bandwidth of 100 Gbps, leading to a very large memory consumption compared to SHIP. Moreover, the memory consumption of this architecture is highly sensitive to the prefix distribution used. For the memory latency, in the worst case, when a collision is detected, two on-chip memory accesses are required, followed by three memory bursts pipelined in a single off-chip DRAM, leading to a total latency of~80~ns. The observed latency of SHIP is 61\% smaller. Finally, no scaling study is presented, making it difficult to appreciate the performance of FlashLook for future applications.

The method proposed in~\cite{scalable_ipv6_vk} uses a tree-based solution that requires $19$ bytes per prefix, which is $78\%$ larger than the proposed SHIP algorithm. Regarding the memory accesses, in the worst case, using a prefix table holding $580$ k prefixes, $22$ memory accesses are required, which is more than twice the number of memory accesses required by SHIP. In terms of latency, their implementation leads to a total latency of $90$ ns for a prefix table holding $580$ k prefixes, that is $2.9 \times$ higher than the proposed SHIP solution. Nevertheless, similar to SHIP, this solution has a logarithmic scaling factor in terms of memory accesses, and scales linearly in terms of memory consumption. 
 
Finally, Tong et al.~\cite{mem_eff} present the CLIPS architecture~\cite{CLIPS} extended to IPv6. Their method uses $27.6$ bytes per prefix, which is about $2.5 \times$ larger than SHIP. The data structure is stored in both on-chip and off-chip memory, but the number of memory accesses per module is not presented by the authors, making it impossible to give an estimate of the memory latency. Finally, the scalability of this architecture has not been discussed by the authors. 

These results show that SHIP reduces the memory consumption over other solutions and decreases the total memory latency to perform a lookup. It also offers a logarithmic scaling factor for the number of memory accesses, and it has a linear memory consumption scaling.

\section{Conclusion}\label{sec:conclusion}

In this paper, SHIP, a scalable and high performance IPv6 lookup algorithm, has been proposed to address current and future application performance requirements. SHIP exploits prefix characteristics to create a shallow and compact data structure. First, two-level prefix grouping leverages the prefix length distribution and prefix density to cluster prefixes into groups that share common characteristics. Then, for each prefix group, a hybrid trie-tree is built. The proposed hybrid trie-tree is tailored to handle local prefix density variations using a density-adaptive trie and a reduced D-Tree leaf structure.

Evaluated with real and synthetic prefix tables holding up to 580 k IPv6 prefixes, SHIP builds a compact data structure that can fit within current on-chip memory, with very low memory lookup latency. Even for the largest prefix table, the memory consumption per prefix is $10.64$ bytes, with a maximum number of $10$ on-chip memory accesses. Moreover, SHIP provides a logarithmic scaling factor in terms of the number of memory accesses and a linear memory consumption scaling. Compared to other approaches, SHIP uses at least 44\% less memory per prefix, while reducing the memory latency by 61\%.

\section*{Acknowledgments}

The authors would like to thank the Natural Sciences and Engineering Research Council of Canada (NSERC), Prompt,
and Ericsson Canada for financial support to this research.

\bibliographystyle{IEEEtran}
\bibliography{Document}

\end{document}